\documentclass[12pt,twoside]{article}   
\usepackage[super,sort,comma]{natbib}
\usepackage{amsfonts}
\usepackage{amsmath}
\usepackage{amssymb}
\usepackage{mathrsfs}
\usepackage{amsbsy}
\usepackage{tensor}
\usepackage{esint}
\usepackage{tikz} 
\usepackage{multirow}
\usepackage{tabularx}
\usepackage{adjustbox}
\usepackage{makecell}
\usepackage{textcomp}
\usepackage{theorem}
\usepackage{epstopdf}
\usepackage[normalem]{ulem}
\usepackage{booktabs,floatrow}
\usepackage{algorithm}
\usepackage{algorithmic}
\usepackage{multirow}

\usepackage{fancyhdr}		

\usepackage[section]{placeins}   %

\usepackage{graphicx}

\makeatletter \renewcommand\@biblabel[1]{$^{#1}$} \makeatother
\newcommand{\cen}[1]{\begin{center} #1 \end{center}}

\usepackage[mathlines]{lineno}

\usepackage{hyperref}
\hypersetup{ colorlinks,
    citecolor=blue,
    filecolor=blue,
    linkcolor=blue,
    urlcolor=blue
}

\usepackage{xcolor}

\definecolor{gray}{rgb}{0.6,0.6,0.6}
\definecolor{red}{rgb}{0.85,0,0}
\definecolor{green}{rgb}{0,0.85,0}
\definecolor{blue}{rgb}{0,0,0.85}
\definecolor{beige}{rgb}{0.92,0.87,0.78}
\usepackage[all]{hypcap}    

                    {$\blacksquare$\vspace*{7pt}} 

\def\f{\frac}

\def\x{\boldsymbol{x}}

\def\z{{\boldsymbol z}}

\def\h{{\boldsymbol h}}

\def\0{{\mathbf{0}}}
\def\h{\mathbf{h}}
\def\bth{{\boldsymbol \theta}}
\def\bphi{{\boldsymbol \phi}}
\def\bpsi{{\boldsymbol \psi}}
\def\bzeta{{\boldsymbol \zeta}}

\def\bmu{{\boldsymbol \mu}}

\def\R{{\mathbb R}}

\def\mB{{\mathcal B}}

\def\Om{\Omega}

\def\etal{{\it et al. }}

\def\PNMAR{P_{\text{\tiny NMAR}}}

\def\pprior{P_{\text{\tiny prior}}}
\def\muprior{\mu_{\text{\tiny prior}}}
\def\muMA{\mu_{\text{\tiny  MA}}}

\def\eref#1{(\ref{#1})}

\newcolumntype{C}{>{\centering\arraybackslash}p{7.5em}}

\begin{document}

\cen{\sf {\Large {\bfseries MGMAR: Metal-Guided Metal Artifact Reduction for X-ray Computed Tomography} \\
\vspace*{10mm}
Hyoung Suk Park$^1$ and Kiwan Jeon$^{1}$} \\
$^1$National Institute for Mathematical Sciences, Daejeon, 34047, Republic of Korea\\
}

\pagenumbering{roman}
\setcounter{page}{1}
\pagestyle{plain}
Correspondence: Kiwan Jeon Email: jeonkiwan@nims.re.kr \\

\begin{abstract}
In X-ray computed tomography (CT), metal artifact reduction (MAR) remains a major challenge because metallic implants violate standard CT forward-model assumptions, producing severe streaking and shadowing artifacts that degrade diagnostic quality. We propose MGMAR, a metal-guided MAR method that explicitly leverages metal-related information throughout the reconstruction pipeline. MGMAR first generates a high-quality prior image by training a conditioned implicit neural representation (INR) using metal-unaffected projections, and then incorporates this prior into a normalized MAR (NMAR) framework for projection completion. To improve robustness under severe metal corruption, we pretrain the encoder-conditioned INR on paired metal-corrupted and artifact-free CT images, thereby embedding data-driven prior knowledge into the INR parameter space. This prior-embedded initialization reduces sensitivity to random initialization and accelerates convergence during measurement-specific refinement. The encoder takes a metal-corrupted reconstruction together with a recursively constructed metal artifact image, enabling the latent field to capture metal-dependent global artifact patterns. After projection completion using the INR prior, we further suppress residual artifacts using a metal-conditioned correction network, where the metal mask modulates intermediate features via adaptive instance normalization to target metal-dependent secondary artifacts while preserving anatomical structures. Experiments on the public AAPM-MAR benchmark demonstrate that MGMAR achieves state-of-the-art performance, attaining an average final score of \(0.89\) on 29 clinical test cases.
\end{abstract}

\newpage     


\newpage

\setlength{\baselineskip}{0.7cm}      

\pagenumbering{arabic}
\setcounter{page}{1}
\pagestyle{fancy}
\section{Introduction}

X-ray computed tomography (CT) is widely utilized across a broad range of clinical applications, including diagnostic imaging, treatment planning, and image-guided interventions \cite{Gjesteby2017,Elnagar2020}. However, the presence of metallic implants, such as hip prostheses, dental fillings, and spinal screws, severely violates the linear attenuation assumption underlying the filtered backprojection (FBP) algorithm \cite{Bracewell1967}, giving rise to a highly nonlinear and ill-posed reconstruction problem \cite{Park2024b}. These violations are primarily attributed to several physical phenomena, including beam hardening, scattering, nonlinear partial volume effects, and photon starvation. Consequently, metal artifacts, typically manifested as streaking and shadowing in reconstructed images, significantly degrade diagnostic quality and hinder reliable clinical interpretation \cite{Gjesteby2017}. As metallic implants are increasingly used in clinical practice, the demand for effective metal artifact reduction (MAR) techniques has become increasingly critical.

Over the past decades, numerous MAR approaches have been proposed, including projection-completion methods  \cite{Abdoli2010,Kalender1987,Lewitt1978,Meyer2010,Park2013,Roeske2003,Zhao2000} and model-based reconstruction methods \cite{DeMan2001,Elbakri2002,Kyriakou2010,Menvielle2005,OSullivan2007,Wang1996}. Projection completion methods typically attempt to restore metal-corrupted sinogram regions through interpolation-based inpainting. To facilitate more accurate interpolation, these methods often normalize the sinogram using a prior image that is obtained via multi-threshold tissue classification (air, soft tissue, bone) \cite{Mehranian2013,Meyer2010,Prell2009}. Although the normalized MAR (NMAR) framework has demonstrated improved robustness, its performance heavily depends on the accuracy and consistency of the prior image \cite{Gjesteby2017,Njiti2024}.

Model-based reconstruction methods aim to estimate energy-independent quantities (e.g., density or monochromatic images) by explicitly modeling the polychromatic X-ray acquisition process, thereby mitigating metal-induced beam hardening effects. Statistical iterative reconstruction techniques update the desired image by maximizing the log-likelihood of the measurements under a polychromatic forward model, often combined with hand-crafted priors such as edge-preserving regularization \cite{Elbakri2002}. However, these methods require detailed prior knowledge of the X-ray spectrum and material-dependent attenuation coefficients, and they typically entail substantial computational burden. A metal-induced beam hardening correction formula has been proposed to mitigate nonlinear beam-hardening effects \cite{Park2015}, yet it remains limited in addressing interactions between metal and other highly attenuating structures such as bone or teeth.

The performance of existing MAR methods has been further enhanced through the integration of deep learning. Projection completion approaches have benefited from learning-based prior image generation, enabling more accurate sinogram restoration and artifact suppression \cite{Yu2021,Zhang2018a,Gjesteby2017}. Deep neural networks have also been employed to directly replace metal-corrupted sinogram segments through data-driven interpolation \cite{Park2018,Peng2017,Zhu2023}. Dual-domain frameworks enhance projection completion by coupling sinogram inpainting with image-domain constraints, thereby improving restoration of the metal-corrupted sinogram \cite{Lin2019,Zhang2020,Zhou2022}.

Implicit neural representations (INRs) have recently emerged as a promising alternative for CT image reconstruction, modeling an image as a continuous function parameterized by a multilayer perceptron (MLP) \cite{Sitzmann2020,Tancik2020}.  This resolution-independent parameterization offers advantages over conventional grid-based representations, particularly for ill-posed CT reconstruction problems \cite{Kim2022,Park2025,Park2026,Shen2022,Zha2022}. Recent studies have incorporated a polychromatic forward model into the training objective, enabling the network to map spatial coordinates (e.g., pixel/voxel positions) to an energy-independent attenuation value \cite{Park2025,Wu2023,Wu2024}. Nevertheless, such polychromatic model-based formulations may not fully address other acquisition effects, such as photon starvation and scatter. 

Motivated by prior works, we propose a metal-guided MAR method (MGMAR) that explicitly incorporates metal-related information throughout the reconstruction pipeline. MGMAR first generates a high-quality prior image by training a conditioned INR using metal-unaffected projections, and subsequently integrates this prior into a normalized MAR (NMAR) framework for projection completion. 

Although INRs provide an effective continuous representation for ill-posed reconstruction problems, their optimization can become unstable when metal-unaffected projections are severely limited by large or multiple metallic objects, leading to reconstructions that vary across different random initializations (see Fig. \ref{fig:random_init}). To improve robustness under such conditions, we embed data-driven prior knowledge directly into the INR parameter space via pretraining-based weight initialization on paired metal-corrupted and artifact-free CT images. This prior-embedded initialization reduces sensitivity to random initialization and provides a reliable starting point for subsequent measurement-specific refinement.

To enable effective knowledge sharing across samples, we adopt an encoder-conditioned INR architecture. While previous encoder-based INR approaches map an input image to a latent code for conditioning a shared MLP~\cite{Chen2019,Mescheder2019,Sitzmann2020}, conventional convolutional encoders are primarily driven by local receptive fields and may struggle to capture globally distributed metal artifact patterns. To address this limitation, we design an artifact-aware latent conditioning strategy in which the encoder takes both a metal-corrupted CT reconstruction and a recursively constructed metal-artifact image as inputs. This design allows the latent field to encode metal-dependent global streaking and shadowing patterns, which is critical for robust prior estimation under severe corruption. The resulting pretrained weights are used to initialize the encoder-conditioned INR, thereby embedding data-driven prior knowledge into the network and improving reconstruction quality with substantially reduced computational cost.

Following projection completion with the INR prior, we further suppress secondary artifacts introduced by imperfect interpolation using a metal-conditioned residual learning network. Specifically, the metal mask modulates intermediate feature maps through an adaptive instance normalization (AdaIN)-based operation, encouraging the network to focus on metal-dependent secondary artifacts while preserving non-metal anatomical structures.

The proposed method was evaluated on the AAPM CT Metal Artifact Reduction (AAPM-MAR) Grand Challenge dataset~\cite{Haneda2025}, which comprises 14{,}000 paired simulated training cases and 29 clinical metal-corrupted test cases assessed using eight clinically relevant image-quality metrics. Our preliminary submission achieved second place in the competition, and the corresponding algorithm is summarized in the benchmark paper~\cite{Haneda2025}. In the previous version, the prior image was generated using a meta-initialization strategy~\cite{Nichol2018,Tancik2021} pretrained solely on ground-truth projections from the training dataset. In this study, we further enhance the framework by introducing the proposed data-driven prior generation strategy and simplifying the overall pipeline by removing a redundant NMAR step. The improved version achieves state-of-the-art performance, attaining an average final score of \(0.89\) on the 29 clinical test cases.

Our contributions include:
(i) A metal-guided MAR framework that explicitly integrates metal-related information throughout the reconstruction pipeline;
(ii) A data-driven weight initialization strategy based on metal artifact-aware latent conditioning, which substantially improves training stability, reconstruction quality, and computational efficiency;
(iii) A metal-embedded residual learning network that effectively suppresses secondary artifacts caused by inaccurate projection-completion;
(iv) State-of-the-art performance on the AAPM-MAR dataset across clinically relevant image-quality metrics.

\section{Method}
We consider a 2D fan-beam CT acquisition geometry with a curved detector, where a fan-shaped X-ray beam traverses the patient's head or body and is recorded at discrete detector bins. Let \(P(\varphi,s)\) denote the metal-corrupted projection data, where \(\varphi \in [0,2\pi)\) represents the projection angle and \(s \in \R\) denotes the detector position along the curved detector. Given \(P\), the conventional FBP reconstructs a CT image \(\mu(\x) = \mB[P](\x)\), where \(\x \in \Om\) denotes a spatial position in the image domain \(\Om\) and \(\mB[\cdot]\) represents the FBP operator \cite{Bracewell1967}.

In the presence of metallic implants in the patient, the projection data \(P\) violates the linear attenuation model assumed by FBP, resulting in pronounced streaking and shadowing artifacts in \(\mu\). These artifacts arise from multiple physical effects, including beam hardening, scattering, and photon starvation. Let \(\mu_*(\x)\) denote the corresponding clean CT image without metal. The objective of this study is to reconstruct \(\mu_*(\x)\) from the metal-corrupted projection data $P$.

We address metal artifacts in the projection domain within the NMAR framework. Let \(\mu_{\text{prior}}\) denote a prior image, and let \(\pprior\) denote its forward projection, defined as
\begin{align}
\pprior(\varphi,s) = \int_{\ell_{\varphi,s}} \mu_{\text{prior}}(\x)\, d\ell_{\x},
\end{align}
where \(\ell_{\varphi,s}\) represents a fan-beam ray connecting the X-ray source at angle \(\varphi\) to the detector position \(s\), and \(d\ell_{\x}\) denotes the line element along the ray. Let \(D \subset \Om\) denote the metal region in the image domain, and let
\(T= \{(\varphi,s)\in[0,2\pi)\times\R:\ \int_{\ell_{\varphi,s}}\chi_D(\x)\, d\ell_{\x} > 0\}\)
denote the metal-affected projection region (i.e., the metal trace). Here, \(\chi_D\) denotes the characteristic function of \(D\), i.e., \(\chi_D(\x)=1\) if \(\x\in D\) and \(0\) otherwise.

Given \(P, \pprior\) and \(T\), the NMAR-corrected projection data, denoted by \(\PNMAR\), is obtained as
\begin{align}\label{PNMAR}
\PNMAR = L\left[\f{P}{\pprior}; T\right]\pprior,
\end{align}
where \(L[\cdot]\) denotes a one-dimensional linear interpolation operator applied along the detector coordinate \(s\) for each fixed projection angle \(\varphi\). It replaces metal-corrupted data in the normalized projection by linearly interpolated values computed from its neighboring metal-unaffected measurements. The NMAR-corrected image is then obtained by the FBP as \(\mu_{\text{\tiny NMAR}}=\mB[\PNMAR]\).

The NMAR reconstruction depends heavily on the quality of the prior image \(\muprior\). Specifically, an accurate prior projection \(\pprior\), obtained by forward projection of \(\muprior\), enables NMAR to compensate primarily for metal-induced corruption by interpolating the normalized data \(P/\pprior\), thereby better preserving anatomical structures outside the metal region \(D\) in the NMAR-corrected image \(\mu_{\text{\tiny NMAR}}\).
In contrast, an inaccurate estimate of \(\pprior\) can induce a large interpolation error in the normalized projection,
\begin{align}\label{interpolation_error}
\big\|L\!\left[\f{P}{\pprior};T\right] - 1\big\|_{T},
\end{align}
which can lead to severe secondary artifacts in the reconstructed image \(\mu_{\text{\tiny NMAR}}\). Here, \(\|\cdot\|_{T}\) denotes a norm over the metal-trace region \(T\).

\begin{figure}[t]
    \centering
    \includegraphics[width=1.0\textwidth]{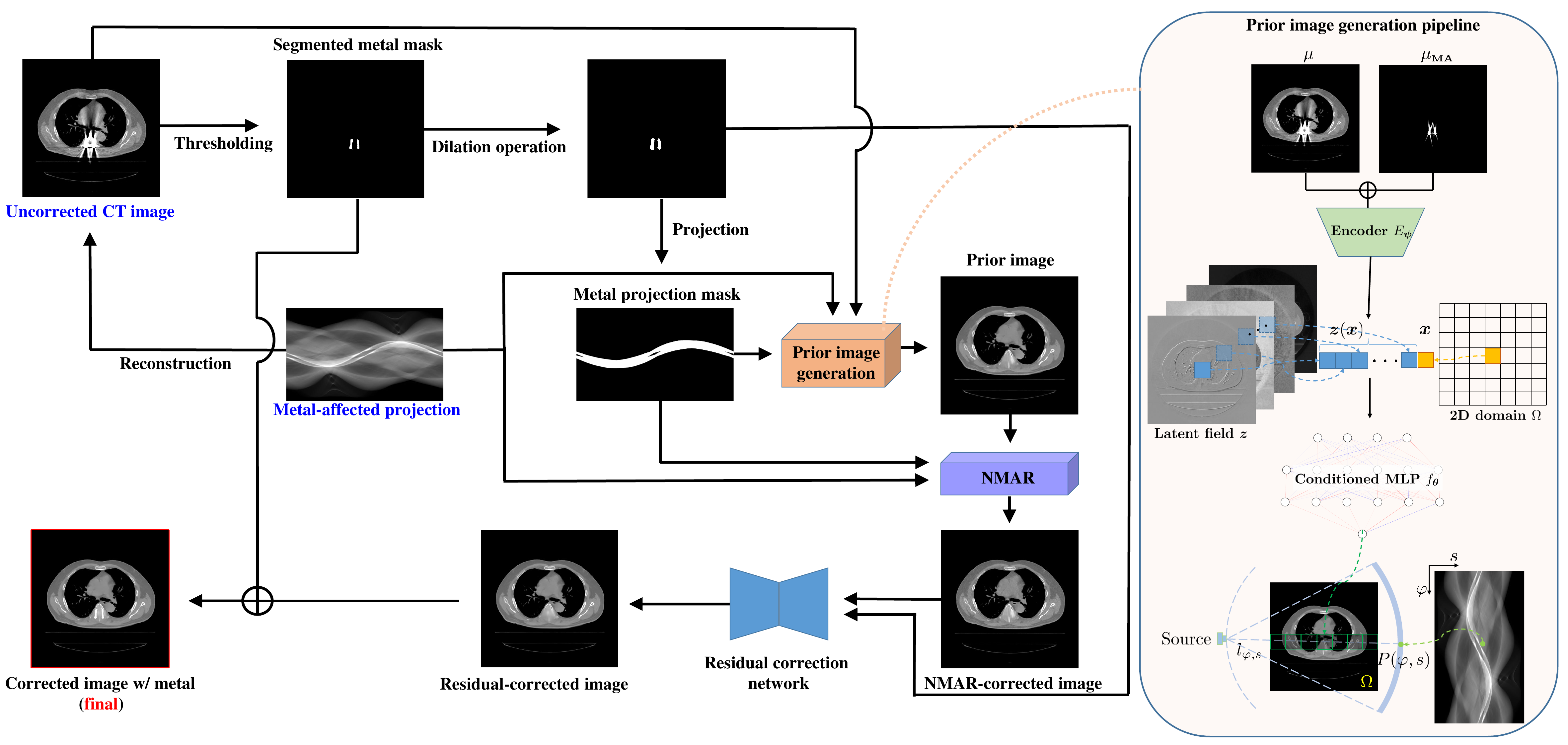}
	\caption{Schematic diagram of the MGMAR framework, which exploits metal-related information throughout the pipeline via a metal artifact-aware latent-conditioned INR for prior image generation and a metal-embedded residual learning network.}
    \label{Overall_framework}
\end{figure}

In the following sections, we describe (i) an effective strategy for generating a high-quality prior image using an INR from metal-unaffected projections, and (ii) a metal-embedded residual correction method to suppress secondary artifacts in \(\mu_{\text{NMAR}}\). An overview of the MGMAR framework is illustrated in Fig.~\ref{Overall_framework}.

\subsection{INR-based prior image generation}
Given a spatial location \(\x\in\Om\), we aim to learn an MLP \(f_{\bth}:\x\mapsto\mu_{\text{prior}}\) parameterized by learnable weights \(\bth\), using metal-unaffected projection measurements \(P(\varphi,s)\) on \(T^c:=([0,2\pi)\times\R)\setminus T\). The network can be trained by minimizing the following loss:
\begin{equation}\label{naive-inr-loss}
   {\mathcal L_{\text{\tiny naive}}(\bth)} = \mathbb E_{(\varphi,s)\in T^c}\left[ \left|P(\varphi, s)- \int_{\ell_{\varphi,s}}  f_{\bth}(\x)\, d\ell_{\x} \right|\right].
\end{equation}

\begin{figure}[ht]
    \centering
    \includegraphics[width=1.0\textwidth]{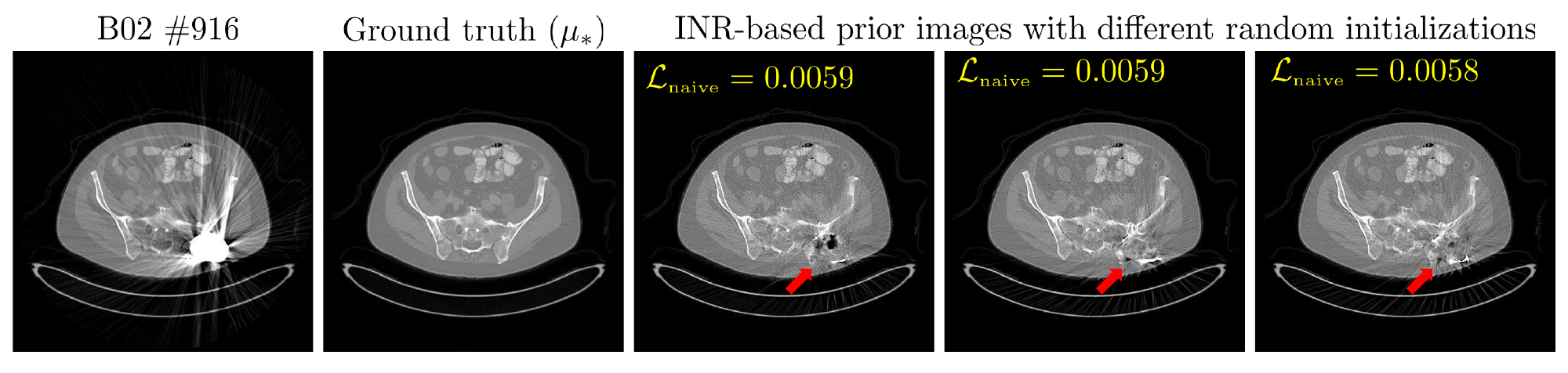}
	\caption{Prior images obtained by an INR with different random initializations. Given the projection data \(P\) from a validation case (B02 \#916), the INR is optimized by minimizing the naive loss \(\mathcal L_{\text{\tiny naive}}\) for \(10{,}000\) iterations (WW/WL \(=1500/-250\) HU for CT images). Although the final loss values are similar, the resulting reconstructions differ substantially.}
    \label{fig:random_init}
\end{figure}

However, despite the reconstruction potential of INRs for the ill-posed problem, directly optimizing \eref{naive-inr-loss} can become unstable when metal-unaffected measurements are severely limited by large and multiple metallic implants. Iteratively updating the weights from random initialization often yields sub-optimal reconstructions that fluctuate across initializations and requires substantial computational cost (see Fig. \ref{fig:random_init}).

\subsubsection{A data-driven weight initialization strategy}
To address this problem, we introduce a data-driven weight initialization that embeds prior knowledge into the MLP weights and serves as a learned initialization instead of random initialization.
The initialization is obtained by pretraining a conditioned INR on a paired training dataset, where the conditioning is provided by an artifact-aware representation extracted by an encoder.
Specifically, let \(E_{\bpsi}:\bmu \mapsto \z\) be a convolutional encoder parameterized by \(\bpsi\), which takes a two-channel input \(\bmu := (\mu,\,\mu_{\text{\tiny MA}})\) and outputs an \(m\)-channel latent field \(\z:\Om\to\R^m\).
Here, \(\mu_{\text{\tiny MA}}\) denotes a metal-artifact image designed to emphasize metallic structures and their associated globally distributed artifact patterns. It serves as an estimate of the quantity \(\mu^*_{\text{\tiny MA}} := \mu - \mu_*\). Since \(\mu_*\) is unavailable during inference, we construct \(\mu_{\text{\tiny MA}}\) using the measured projection data \(P\) through a simple recursive procedure described later.

We pretrain the conditioned INR by jointly learning \((\bth,\bpsi)\) from the paired training dataset \(\mathcal S_{\text{\tiny init}}=\{(\bmu^i, \mu_*^i)\}_{i=1}^N\) via
\begin{align}\label{data-drive-weight}
\mathcal L_{\text{\tiny init}}(\bth,\bpsi;\mathcal S_{\text{\tiny init}})
= \f{1}{N}\sum_{i=1}^N\mathbb E_{\x\in\Om}\!\left[\left|f_{\bth}\!\left(\x,\, E_{\bpsi}(\bmu^i)(\x)\right)-\mu^{i}_*(\x)\right|\right]
\end{align}

\begin{figure}[ht]
    \centering
    \includegraphics[width=1.0\textwidth]{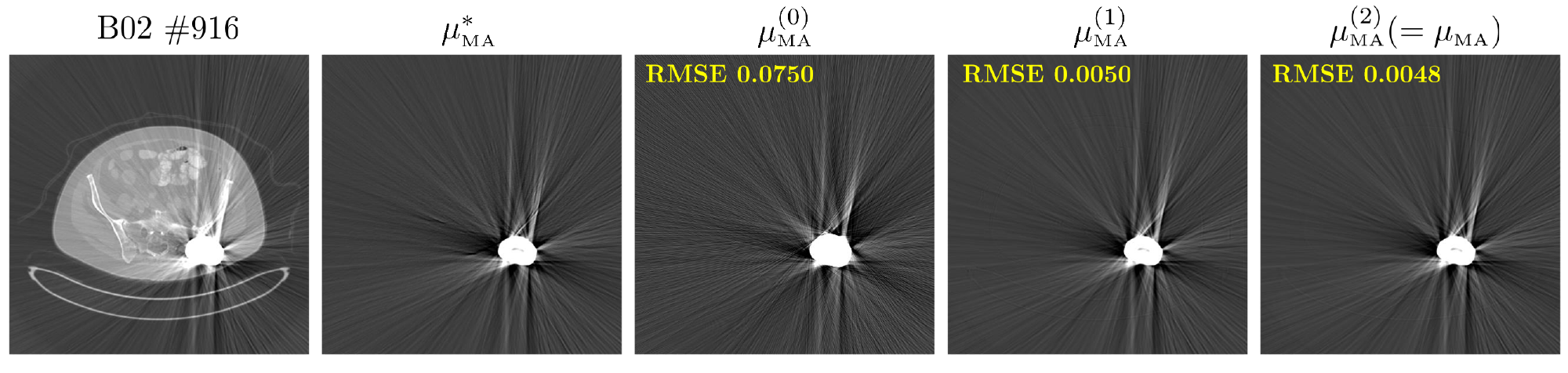}
	\caption{Constructed metal-artifact images \(\mu_{\text{\tiny MA}}^{(k)}\) across recursion steps (\(k=0,1,2\)). The RMSE with respect to the target  \(\mu_{\text{\tiny MA}}^{*}\) is reported at the top of each panel (WW/WL \(=2000/-500\)).}
    \label{fig:mu_MA}
\end{figure}

\paragraph{Recursive construction of metal-artifact image \(\mu_{\text{\tiny MA}}\).}
We estimate \(\mu^*_{\text{\tiny MA}}\) in a recursive manner. First, we set the initial estimate \(\mu_{\text{\tiny MA}}^{(0)}\) as
\[
\mu_{\text{\tiny MA}}^{(0)} := \mB\!\left[\chi_T P\right],
\]
where \(\chi_T\) is the metal projection mask. We then update \(\mu_{\text{\tiny MA}}^{(k)}\) by the recursion
\[
\mu_{\text{\tiny MA}}^{(k+1)} := \mB\!\left[P-P_{\text{\tiny NMAR}}^{(k)}\right], \qquad k=0,1,\ldots,K-1,
\]
where \(P_{\text{\tiny NMAR}}^{(k)}\) is the NMAR-corrected projection computed using a prior projection \(P_{\text{\tiny prior}}^{(k)}\) as
\begin{align}
P_{\text{\tiny NMAR}}^{(k)} = L\left[\f{P}{P_{\text{\tiny prior}}^{(k)}}; T\right]P_{\text{\tiny prior}}^{(k)}.
\end{align}
The prior projection \(P_{\text{\tiny prior}}^{(k)}\) is obtained from the conditioned INR as
\[
P_{\text{\tiny prior}}^{(k)}(\varphi,s) = \int_{\ell_{\varphi,s}} f_{\bth^{(k)}}\!\left(\x,\z^{(k)}(\x)\right)\, d\ell_{\x}, 
\]
where $\z^{(k)}(\x)=E_{\bpsi^{(k)}}(\bmu^{(k)})(\x)$ and  \(\bmu^{(k)} := (\mu,\,\mu_{\text{\tiny MA}}^{(k)})\).
In each recursion step, the weights \((\bth^{(k)},\bpsi^{(k)})\) are pretrained by minimizing \(\mathcal L_{\text{\tiny init}}\) in \eref{data-drive-weight} on a training set constructed using the current artifact estimate, denoted by \(\mathcal S_{\text{\tiny init}}^{(k)}=\{(\bmu^{i,(k)}, \mu_*^i)\}_{i=1}^N\), where \(\bmu^{i,(k)}:=(\mu^i,\,\mu_{\text{\tiny MA}}^{i,(k)})\).
Finally, we set \(\mu_{\text{\tiny MA}} := \mu_{\text{\tiny MA}}^{(K)}\) for a suitably chosen \(K\). Fig.~\ref{fig:mu_MA} shows an example of the constructed metal-artifact images \(\mu_{\text{\tiny MA}}^{(k)}\) at successive recursion steps (\(k=0,1,2\)). As the recursion proceeds, \(\mu_{\text{\tiny MA}}^{(k)}\) progressively approaches the target artifact-only image \(\mu_{\text{\tiny MA}}^{*}\) in terms of RMSE.

\subsubsection{Measurement-specific adaptation of the INR} 
Starting from the data-driven initialization \(\bth_0:=\bth^{(K)}\) obtained in \eref{data-drive-weight}, we further adapt the INR to each measurement \(P\) by enforcing data fidelity on the metal-unaffected region \(T^c\), subject to structural constraints derived from the initial weight $\bth_0$.  Specifically, for a given encoder input \(\bmu=(\mu,\mu_{\text{\tiny MA}})\), we refine \(f_{\bth}\) by minimizing
\begin{equation}\label{proposed-inr-loss}
   {\mathcal L_{\text{\tiny fid}}(\bth)} = \mathbb E_{(\varphi,s)\in T^c}\left[ \left|P(\varphi, s)- \int_{\ell_{\varphi,s}}  f_{\bth}(\x,\z(\x))\, d\ell_{\x} \right|\right] + \lambda \mathbb E_{\x\in \Om}\left[ \left|f_{\bth}(\x,\z(\x))-f_{\bth_0}(\x,\z(\x))\right|\right],
\end{equation}
where \(\z=E_{\bpsi^{(K)}}(\bmu)\) is the latent field extracted by the pretrained encoder in \eref{data-drive-weight}, and $\lambda>0$ is the regularization parameter. Finally, we obtain the prior image by evaluating the refined INR at each spatial location, i.e.,
\(\mu_{\text{prior}}(\x) := f_{\bth}(\x,\z(\x))\) for all \(\x\in\Om\). The overall inference procedure for generating the prior image using the data-driven initialization is summarized in Algorithm~\ref{alg:inr_prior_inference}.

\begin{algorithm}[t]
\caption{Prior image generation with data-driven initialization for NMAR}
\label{alg:inr_prior_inference}
\begin{algorithmic}[1]
\REQUIRE Metal-corrupted projection \(P(\varphi,s)\); metal trace \(T\); metal trace mask \(\chi_T\); FBP reconstruction \(\mu=\mB[P]\);
pretrained encoders \(\{E_{\bpsi^{(k)}}\}_{k=0}^{K}\); pretrained INRs \(\{f_{\bth^{(k)}}\}_{k=0}^{K}\);
recursion steps \(K\); refinement iterations \(N_{\text{iter}}\).
\ENSURE Prior image \(\mu_{\text{prior}}\).

\STATE \textbf{(Construction of \(\mu_{\text{\tiny MA}}\))} \; \(\mu_{\text{\tiny MA}}^{(0)} \leftarrow \mB[\chi_T P]\).

\FOR{\(k=0,1,\ldots,K-1\)}
    \STATE \(\bmu^{(k)} \leftarrow (\mu,\mu_{\text{\tiny MA}}^{(k)})\), \quad
    \(\z^{(k)} \leftarrow E_{\bpsi^{(k)}}(\bmu^{(k)})\).
    \STATE \(P_{\text{\tiny prior}}^{(k)}(\varphi,s) \leftarrow \int_{\ell_{\varphi,s}} f_{\bth^{(k)}}\!\left(\x,\z^{(k)}(\x)\right)\, d\ell_{\x}\), \quad \(\forall (\varphi,s)\in[0,2\pi)\times\R\).
    \STATE \(P_{\text{\tiny NMAR}}^{(k)} \leftarrow L\!\left[\f{P}{P_{\text{\tiny prior}}^{(k)}};\,T\right]P_{\text{\tiny prior}}^{(k)}\).
    \STATE \(\mu_{\text{\tiny MA}}^{(k+1)} \leftarrow \mB\!\left[P-P_{\text{\tiny NMAR}}^{(k)}\right]\).
\ENDFOR

\STATE \(\mu_{\text{\tiny MA}} \leftarrow \mu_{\text{\tiny MA}}^{(K)}\), \quad
\(\bmu \leftarrow (\mu,\mu_{\text{\tiny MA}})\), \quad
\(\z \leftarrow E_{\bpsi^{(K)}}(\bmu)\), \quad \(\bth_0 \leftarrow \bth^{(K)}\).

\STATE \textbf{(Refinement of INR on \(T^c\))}\;
\(\bth \leftarrow \textsc{Opt}\!\left(\bth_0,\; \mathcal L_{\text{\tiny fid}}(\bth),\; N_{\text{iter}}\right)\).

\STATE \(\mu_{\text{prior}}(\x) \leftarrow f_{\bth}(\x,\z(\x))\), \quad \(\forall \x\in\Om\).
\RETURN \(\mu_{\text{prior}}\).
\end{algorithmic}
\end{algorithm}

\subsection{Metal-embedded residual learning}
To suppress the remaining artifacts in the NMAR-corrected image \(\mu_{\text{\tiny NMAR}}\), we adopt an image-domain residual learning strategy.
We denote the residual artifacts by \(r=\mu_{\text{\tiny NMAR}}-\mu_*\).
Motivated by the observation that these artifacts are prominent around metallic implants, we condition the residual network on metal features.
Specifically, let \(G_{\bphi}:(\mu_{\text{\tiny NMAR}}; \chi_D)\mapsto r\) be a residual network parameterized by \(\bphi\), where \(\chi_D\) denotes the metal mask corresponding to the given reconstruction \(\mu_{\text{\tiny NMAR}}\).
We train \(G_{\bphi}\) on the paired dataset \(\mathcal S_{\text{\tiny res}}=\{(\mu_{\text{\tiny NMAR}}^{i}, r^i)\}_{i=1}^N\) by minimizing
\begin{align}\label{residual-loss}
\mathcal L_{\text{\tiny res}}(\bphi;\mathcal S_{\text{\tiny res}})
= \f{1}{N}\sum_{i=1}^N
\mathbb E_{\x\in\Om}\!\left[
\left|
G_{\bphi}\!\left(\mu^{i}_{\text{\tiny NMAR}}; \chi_D^{i} \right)(\x)
-r^{i}(\x)
\right|
\right].
\end{align}
Here, the metal mask \(\chi_D^{i}\) is fed into \(G_{\bphi}\) through an AdaIN-based conditioning mechanism \cite{Huang2017,Karras2019}.

\paragraph{AdaIN-based metal conditioning.}
We first assume that \(G_{\bphi}\) is implemented as a multi-scale convolutional encoder-decoder. Let \(\h:\Om\to\R^{J_\h}\) denote an intermediate feature field generated within \(G_{\bphi}\) during the forward pass, where \(\h(\x)=[h_1(\x),\ldots,h_{J_\h}(\x)]^\top\) and \(h_j:\Om\to\R\) denotes its \(j\)-th channel. Here, \(J_\h\) is the number of feature channels of $\h$. AdaIN modulates each channel by matching its spatial mean and standard deviation to target statistics predicted from the metal mask \(\chi_D\).
Formally, given $\h$, the AdaIN-modulated channel-wise feature field is defined by
\begin{align}
\Gamma\!\left(h_j;\alpha_j,\beta_j\right)(\x)
=
\f{\beta_j}{\sigma_{\x\in\Om}\!\left[h_j\right]}
\left(h_j(\x)-\mathbb E_{\x\in\Om}\!\left[h_j(\x)\right]\mathbf 1(\x)\right)
+\alpha_j\mathbf 1(\x),
\end{align}
where \(\mathbf 1:\Om\to\R\) denotes the constant-one function, and \(\alpha_j\) and \(\beta_j\) are channel-wise affine parameters.
These parameters are produced by a network \(F_{\bzeta}\) conditioned on the metal mask \(\chi_D\):
\begin{align}
F_{\bzeta}:\chi_D \mapsto \big\{(\alpha_j,\beta_j)\big\}_{j=1}^{J_\h}.
\end{align}
The network \(F_{\bzeta}\) is jointly learned with \(G_{\bphi}\), enabling metal-dependent modulation of multi-scale features. A schematic of the residual network with AdaIN-based conditioning is provided in the AAPM-MAR benchmark paper~\cite{Haneda2025} (Fig.~5(b)), where an earlier version of our MAR method submitted to the AAPM-MAR challenge is summarized.
 
\section{Experiments}
\subsection{Experimental Settings}

\paragraph{Datasets.}
We evaluate our method on the public AAPM-MAR benchmark dataset. The benchmark contains \(14{,}000\) simulated 2D CT datasets generated using a hybrid CT simulation pipeline, which combines real clinical head and body CT images with virtual fractal-shaped metal objects inserted at random locations in soft tissue or bone. The simulated metal materials include amalgam, stainless steel, copper, cobalt, and titanium. Each dataset provides paired data in both projection and image domains, including sinograms with and without metal, FBP-reconstructed images with and without metal artifacts, and a binary metal mask. The reconstructed images are provided on a \(512\times512\) grid with a field of view (FOV) of \(400\)~mm for body data and \(220.16\)~mm for head data. We split the \(14{,}000\) simulated datasets into training and validation sets with a \(9{:}1\) ratio, resulting in \(12{,}600\) training datasets and \(1{,}400\) validation datasets.

For testing, the benchmark provides \(29\) clinical metal-corrupted datasets in both sinogram and image domains, without ground-truth sinograms or images available. These datasets cover a wide range of clinically relevant scenarios, including small implants (e.g., surgical clips, fiducial markers, and dental fillings), medium-sized implants (e.g., pacemakers and spinal fixation devices), and large joint replacements (e.g., shoulder and hip prostheses).

\paragraph{Metrics and Baselines.}
For the validation datasets, where artifact-free references are available, we compute conventional image-quality metrics including RMSE, PSNR, and SSIM. For the test datasets, evaluation follows the AAPM-MAR benchmark protocol using eight clinically relevant scoring metrics: CT number (CTN) accuracy, noise, image sharpness, streak amplitude, structural integrity (SSIM), metal integrity, bone integrity, and proton beam range (PBR) influence. Detailed definitions of these metrics are provided in \cite{Haneda2025,Peters2025}.
Each metric is normalized to a score in \([0,4]\), where \(0\) corresponds to the ground-truth reference and \(4\) corresponds to the uncorrected image. The final benchmark score is computed as the average of the eight normalized metrics across all \(29\) test cases.  All validation metrics and benchmark scores are computed using the official Python code released in the challenge's GitHub repository \cite{AAPM_CTMAR_Benchmark}.

We compare our method with representative baselines reported in the benchmark paper~\cite{Haneda2025}, including the top three approaches from the final leaderboard and the conventional NMAR method \cite{Meyer2010}. The first-ranked method adopts a dual-learning framework that enforces cycle consistency between projections and reconstructed images while explicitly modeling physical effects such as photon starvation, beam hardening, noise, and scatter. It employs multiple network architectures, including U-Net, CNN, and MLP, within the framework. The third-ranked method employs a dual-domain framework that first performs metal segmentation in both the projection and image domains, and then alternates between sinogram inpainting and image-domain artifact reduction to progressively refine the reconstructed image. It employs a Swin Transformer-based U-Net and a ResNet within the framework. The second-ranked method corresponds to our preliminary submission. In this work, we further improve performance by introducing the proposed data-driven prior image generation and simplifying the overall pipeline by removing a redundant NMAR step previously applied at the end of the procedure, while keeping the same  metal segmentation protocol used in our prior work.

\paragraph{Implementation details.} The implementation is performed using the PyTorch framework \cite{Paszke2017}. The MLP network $f_{\bth}$ consists of $4$ hidden fully connected layers, each with $256$ units and ReLU activation \cite{Nair2010}. For the conditioning encoder \(E_{\bpsi}\), we adopt an EDSR-baseline architecture \cite{Lim2017} with \(16\) residual blocks and \(64\) feature channels. We use a no-upsampling configuration so that the encoder produces a latent feature map at the same spatial resolution as the input image. This enables pixel-wise conditioning of the INR, i.e., \(\z(\x)=E_{\bpsi}(\bmu)(\x)\).

At test time, the artifact image \(\mu_{\text{\tiny MA}}\) is constructed via the recursive procedure with \(K=2\) recursion steps (Algorithm~\ref{alg:inr_prior_inference}). We pretrain each stage-specific conditioned INR \((E_{\bpsi^{(k)}},f_{\bth^{(k)}})\), for \(k=0,1,\ldots,K\), on the paired dataset \(\mathcal S^{(k)}_{\text{\tiny init}}\) by minimizing \(\mathcal L_{\text{\tiny cond}}\) in \eref{data-drive-weight} using a learning rate of \(10^{-4}\) for \(1{,}000\) epochs. For each measurement \(P\), the globally initialized weight \(\bth_0=\bth^{(K)}\) is further optimized by solving the data-fidelity problem in \eref{proposed-inr-loss} with a learning rate of \(5\times10^{-6}\) for \(N_{\text{iter}}=1000\) iterations and the regularization parameter \(\lambda=10\).  During this refinement, the encoder is fixed to \(E_{\bpsi^{(K)}}\).

The residual network \(G_{\bphi}\) is implemented as a multi-scale convolutional encoder--decoder based on deep convolutional framelets~\cite{Ye2018}. At each scale, the network consists of two consecutive \(4\times 4\) convolutional layers, each followed by an AdaIN layer and a LeakyReLU activation \cite{Maas2013}. It employs 2-D Haar wavelet decomposition and reconstruction \cite{Mallat2002} to perform downsampling and upsampling, respectively.

The AdaIN parameter prediction network \(F_{\bzeta}\) follows the branch network architecture \cite{Park2024}. It takes the conditioning metal mask \(\chi_D\) as input and jointly predicts the affine parameters for all AdaIN layers in \(G_{\bphi}\) in a single forward pass. Specifically, \(F_{\bzeta}\) consists of a shared convolutional trunk with \(4\) convolution layers, followed by a flatten operation and two fully connected layers. From the final latent vector, \(F_{\bzeta}\) employs parallel output heads to produce the channel-wise shift and scale parameters $\big\{(\alpha_j,\beta_j)\big\}_{j=1}^{J_\h}$ for each AdaIN-modulated feature field \(\h\) in \(G_{\bphi}\), yielding an output of dimension \(2J_\h\) per feature field.

The residual learning is performed in a patch-by-patch manner. In each iteration, we randomly extract \(16\) paired patches of size \(128\times128\) from the NMAR-corrected image \(\mu_{\text{\tiny NMAR}}\) and the corresponding metal mask 
\(\chi_D\). These patch pairs are used to train \(G_{\bphi}\) with a learning rate of \(10^{-4}\) for \(200\) epochs.

Finally, to estimate the metal region \(D\) in the uncorrected CT image \(\mu\), we employ a simple thresholding method.
Specifically, a threshold value of \(0.12\) is used for the validation datasets, while the metal masks for the test datasets are further manually refined. 
Before forward projection to compute the metal trace, the estimated metal mask is dilated using a disk-shaped structuring element of radius \(2\) \cite{Serra1983}.

\subsection{Results}
\begin{figure}[ht]
    \centering
    \includegraphics[width=1.0\textwidth]{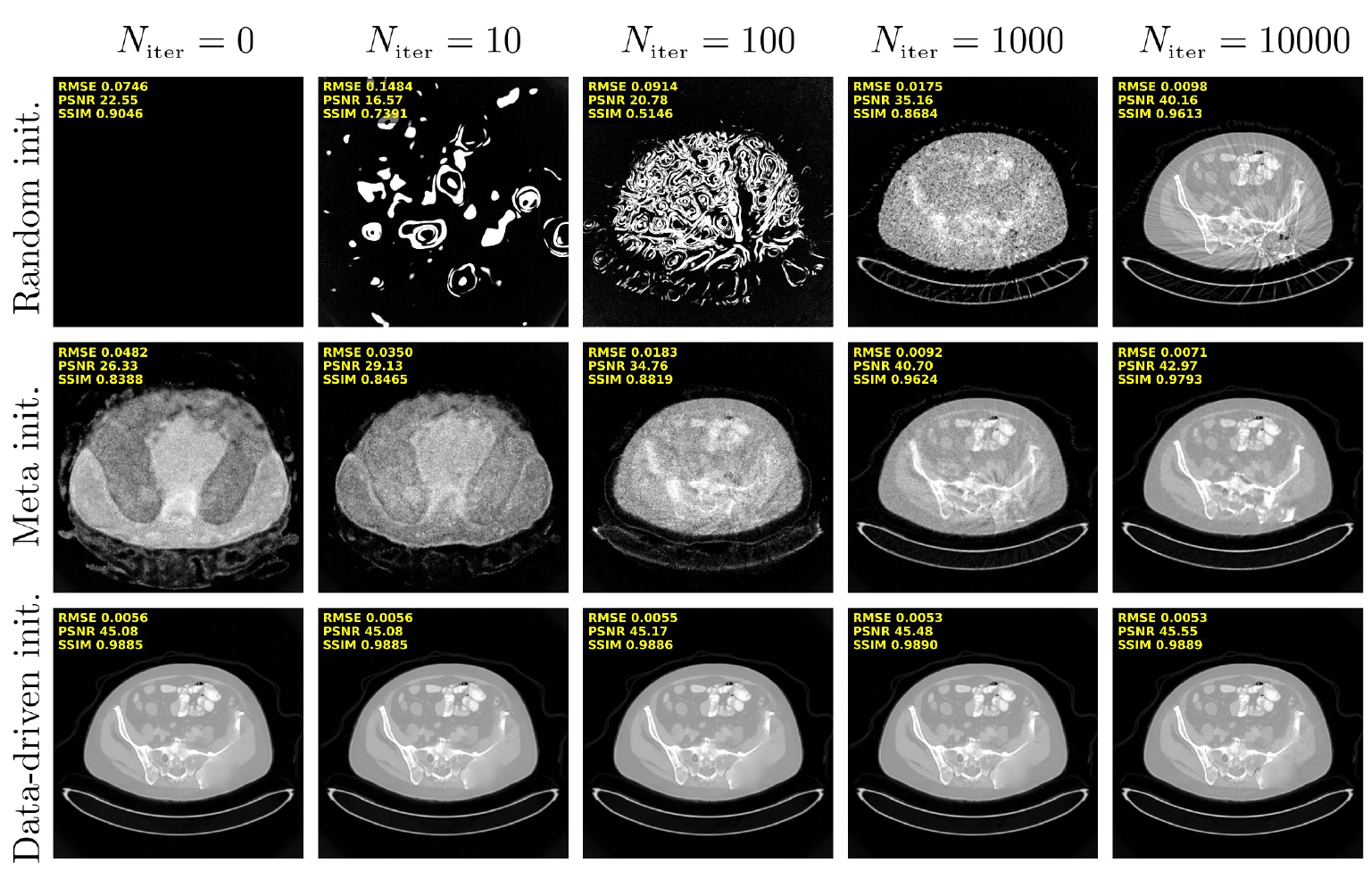}
	\caption{Comparison of prior images obtained by INRs initialized with random, meta, and the proposed data-driven weights across the number of iterations \(N_{\text{\tiny iter}}\) (WW/WL = 1500/-250 HU for CT images).}
	\label{fig:prior_image_generation}
\end{figure}

\paragraph{Effect of data-driven initialization for prior image generation.}
Fig.~\ref{fig:prior_image_generation} compares prior images obtained by INRs initialized with random, meta, and proposed data-driven weights across the number of iterations (\(N_{\text{\tiny iter}}\)).
In this study, the meta initialization is obtained by pretraining the network on \(5{,}000\) ground-truth projection datasets randomly sampled from the training set, using a learning rate of \(5\times10^{-4}\). As shown in Fig.~\ref{fig:prior_image_generation}, compared to random and meta initializations, the proposed data-driven initialization produces high-quality prior reconstructions even without network training (i.e., \(N_{\text{\tiny iter}}=0\)), achieving near-optimal RMSE, PSNR, and SSIM while preserving anatomical structures in metal-affected regions with minimal degradation. Subsequent iterations lead to gradual improvements in prior quality, with performance saturating around \(N_{\text{iter}}=1000\). 

\begin{figure}[ht]
    \centering
    \includegraphics[width=1.0\textwidth]{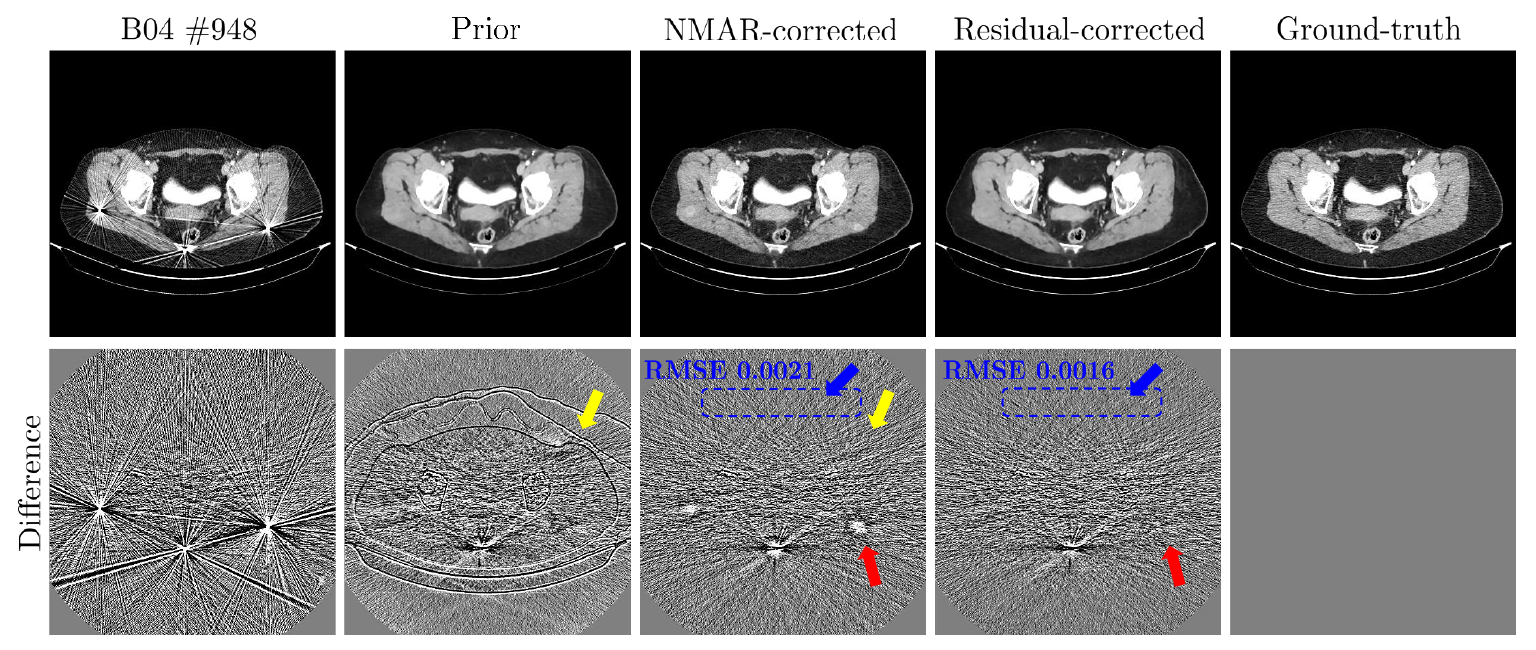}
	\caption{Visual comparison of prior, NMAR-corrected, and residual-corrected images on a validation case (B04 \#948) (WW/WL = 300/0 HU for CT images, WW/WL = 50/0 HU for difference images).}
	\label{fig:nmar-res}
\end{figure}

\paragraph{Effects of NMAR and residual correction.}
Fig.~\ref{fig:nmar-res} illustrates the impact of NMAR and residual correction. While the INR-based prior image exhibits high-quality reconstruction, it fails to fully preserve structural sharpness, resulting in slightly oversmoothed boundaries (yellow arrows). The NMAR-corrected image better preserves the overall structural sharpness, as it modifies only the metal-trace region of the measured projection data while retaining the original measurements elsewhere. However, it suffers from secondary artifacts caused by inaccurate interpolation, particularly in the vicinity of metallic objects (red arrows). The residual learning effectively suppresses these secondary artifacts and mitigates the additional streaking artifacts (blue arrows) in the NMAR-corrected image. For quantitative analysis, RMSE is computed within the blue rectangular region marked in the figure. The RMSE decreases from 0.0021 to 0.0016.

Table~\ref{tab:validation} summarizes the quantitative performance of each stage (i.e., prior-image generation, NMAR, and residual correction) on the entire validation set, along with the average runtime per case. All runtimes are measured on a system equipped with an Intel Xeon Gold 6226R CPU (2.90~GHz) and an NVIDIA RTX 3090 GPU (24~GB). Each stage consistently improves RMSE, PSNR, and SSIM. In terms of computational cost, prior-image generation dominates the total runtime, as it involves the recursive construction of the metal-artifact image \(\mu_{\text{\tiny MA}}\) and iterative refinement of the MLP network \(f_{\bth}\) (i.e., \(N_{\text{\tiny iter}}=1000\)). In contrast, NMAR and residual correction introduce only minor overhead.

\begin{table}[]
\caption{Quantitative results of each stage on the validation set, along with the average runtime per case. The quantitative results are reported as mean \(\pm\) standard deviation.}
\label{tab:validation}
\renewcommand{\arraystretch}{1.3}
\begin{tabular}{lcccc}
                              	& RMSE $\downarrow$ 				& PSNR $\uparrow$ 			& SSIM $~\uparrow$ 				& Runtime (s) \\ \hline
Uncorrected                   	& 0.0143 $\pm$ 0.0059   			& 37.44 $\pm$ 3.01    		& 0.9307 $\pm$ 0.0295  			& -           \\ \hline
Prior image                   	&                   				&                 			&                  				&             \\
$N_{\text{\tiny iter}}=0$     	& 0.0045 $\pm$ 0.0029   			& 47.82 $\pm$ 3.56    		& 0.9857 $\pm$ 0.0161  			& 6           \\
$N_{\text{\tiny iter}}=1000$ 	& 0.0042 $\pm$ 0.0024   			& 48.36 $\pm$ 3.52    		& 0.9860 $\pm$ 0.0162  			& 41          \\ \hline
NMAR-corrected                	& 0.0037 $\pm$ 0.0017   			& 49.24 $\pm$ 3.43    		& 0.9869 $\pm$ 0.0113  			& 2           \\ \hline
Residual-corrected            	& \textbf{0.0032 $\pm$ 0.0016}	& \textbf{50.74 $\pm$ 3.60}	& \textbf{0.9904 $\pm$ 0.0097}  	& 1          
\end{tabular}
\renewcommand{\arraystretch}{1.0}
\end{table}

\begin{figure}[ht]
    \centering
    \includegraphics[width=0.8\textwidth]{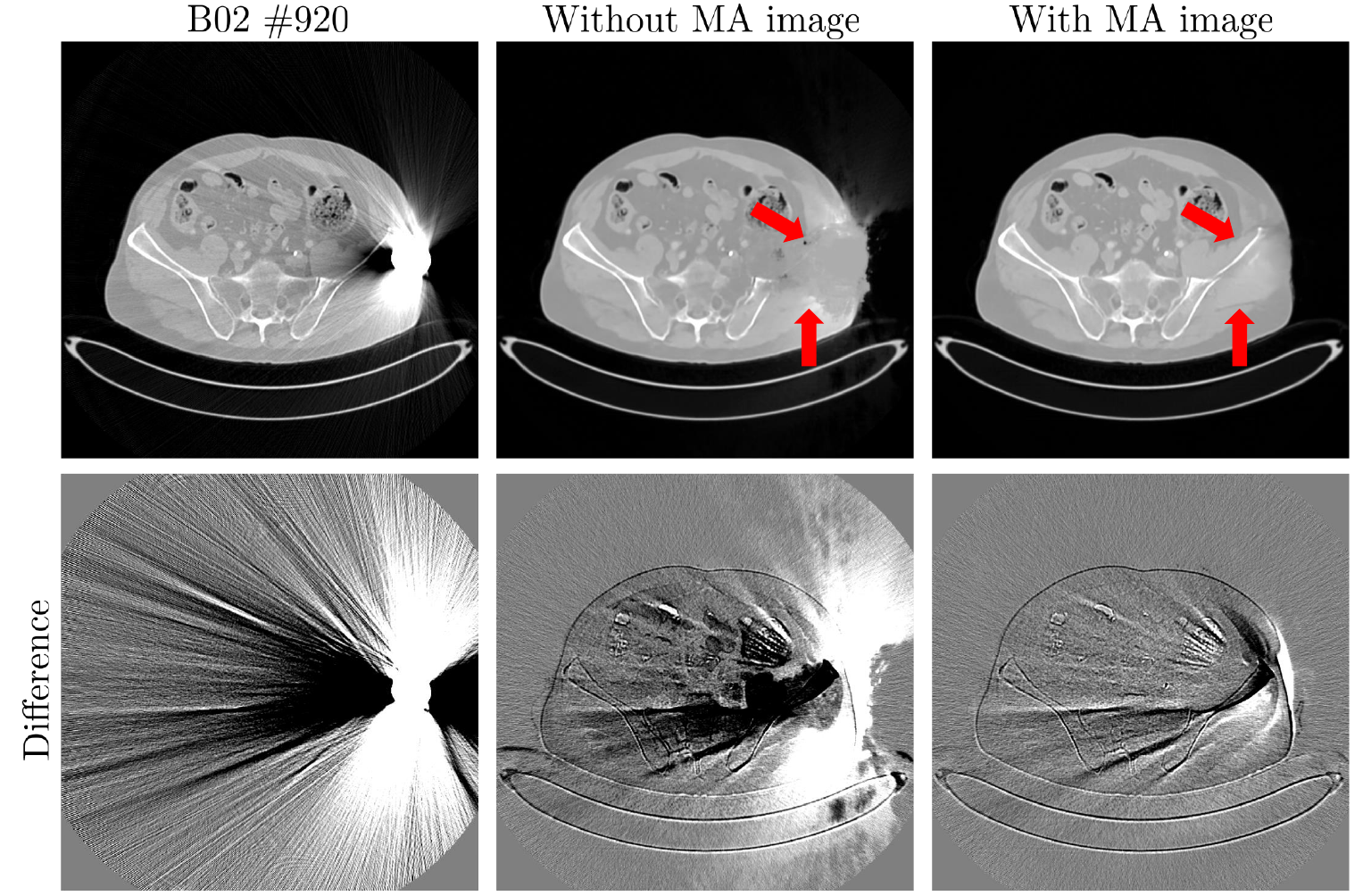}
	\caption{Effect of the metal-artifact image \(\mu_{\text{\tiny MA}}\) for prior image generation (WW/WL = 1500/-250 HU for CT images, WW/WL = 200/0 HU for difference images).} 
    \label{fig:MA_effects}
\end{figure}

\paragraph{Ablation study.}
Fig.~\ref{fig:MA_effects} investigates the effect of incorporating the metal-artifact image \(\mu_{\text{\tiny MA}}\) into the encoder input for prior image generation. The second and third panels show prior images produced by the pretrained network \(f_{\bth_0}\) under two pretraining configurations: the data-driven initial weight \(\bth_0\) is obtained from a model whose encoder is trained with \(\mu\) as a single-channel input, and from a model whose encoder is trained with the two-channel input \((\mu,\mu_{\text{\tiny MA}})\), respectively. As shown in Fig.~\ref{fig:MA_effects}, incorporating \(\mu_{\text{\tiny MA}}\) substantially improves prior estimation under severe metal corruption. It better preserves anatomical structures while suppressing streaking and shadowing artifacts (red arrows).

\begin{figure}[ht]
    \centering
    \includegraphics[width=1.0\textwidth]{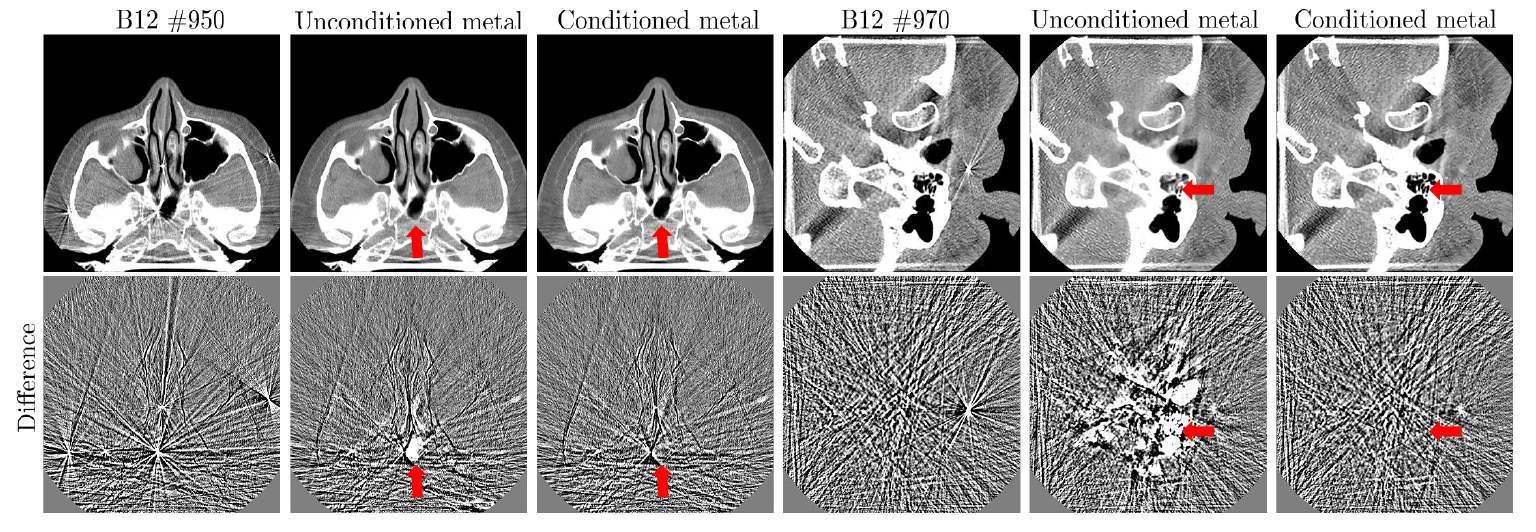}
	\caption{Effect of metal conditioning in residual learning (WW/WL = 1000/0 HU for CT images, WW/WL = 200/0 HU for difference images).} 
    \label{fig:res_effects}
\end{figure}

We also examine the impact of metal conditioning in the residual learning stage. The corresponding results for two different validation cases are shown in Fig.~\ref{fig:res_effects}. For each case, the middle and right panels show residual-corrected results obtained by the pretrained network without and with AdaIN-based metal conditioning, respectively. Without metal conditioning, the residual network tends to introduce unintended distortions of anatomical structures (red arrows). In contrast, metal conditioning guides the network to focus the correction on metal-dependent secondary artifacts while better preserving other anatomical details.


Table~\ref{tab:ablation-table} summarizes the quantitative results of the ablation study on the entire validation set. Incorporating the metal-artifact image \(\mu_{\text{\tiny MA}}\) into the encoder input yields a significant improvement in prior image generation, consistently reducing RMSE and increasing PSNR and SSIM. In contrast, the gain from metal conditioning in the residual learning stage is relatively modest. This may be because the residual artifacts remaining after NMAR are small in magnitude, leaving limited room for further improvement in quantitative metrics.

\begin{table}[]
\caption{Ablation study on the effects of the metal-artifact image \(\mu_{\text{\tiny MA}}\) in prior image generation and metal conditioning in the residual learning stage. The quantitative results are reported on the validation set.}
\label{tab:ablation-table}
\renewcommand{\arraystretch}{1.3}
\begin{tabular}{lccc}
                         		& RMSE $\downarrow$ 				& PSNR $\uparrow$ 				& SSIM $~\uparrow$ 					\\ \hline
Data-driven initialization~~~~  &                   				&                 				&                  					\\
without $\muMA$                 & 0.0053 $\pm$ 0.0031   			& 46.46 $\pm$ 3.52    			& 0.9835 $\pm$ 0.0179 				\\
with $\muMA$                	& \textbf{0.0045 $\pm$ 0.0029}	& \textbf{47.82 $\pm$ 3.56}    	& \textbf{0.9857 $\pm$ 0.0161}  	\\ \hline
Residual correction        		&                   				&                 				&                  					\\
without $\chi_D$                & 0.0033 $\pm$ 0.0017   			& 50.67 $\pm$ 3.64    			& 0.9903 $\pm$ 0.0102  				\\
with $\chi_D$                 	& \textbf{0.0032 $\pm$ 0.0016}	& \textbf{50.74 $\pm$ 3.60}   	& \textbf{0.9904 $\pm$ 0.0097} 
\end{tabular}
\renewcommand{\arraystretch}{1.0}
\end{table}

\paragraph{Results on the test dataset.}
Table~\ref{tab:test_score} reports the final benchmark score and the average value of each evaluation metric over all 29 test cases for the top three methods (ranked first, second, and third) in the AAPM-MAR challenge, conventional NMAR, and MGMAR. MGMAR  achieves the best final score of 0.89 (or 0.8890 when rounded to four decimal places) among the compared MAR approaches. It attains the best performance on four evaluation metrics: CTN accuracy, noise, SSIM, and PBR influence. Compared to our preliminary submission (the second-ranked method), the enhanced version of our method improves most metrics, except for sharpness and metal integrity. The reduction in sharpness is attributed to the final residual learning stage, which suppresses secondary artifacts including streaks and noise but may also attenuate fine structures, resulting in a lower sharpness score. Moreover, although the same metal mask is used as in the preliminary method, the metal integrity score differs (1.09 for the previous version vs.\ 1.13 for the enhanced version). This is because the benchmark evaluates metal integrity based on voxels whose intensities exceed a predefined threshold. We observed that, in many cases, high-intensity values above this threshold in non-metal regions are counted as metal, thereby affecting the reported metal integrity.

\begin{table}[]
\caption{Final benchmark score and average values of each evaluation metric over all 29 test cases for the selected methods ranked first, second, and third in the AAPM-MAR challenge, traditional NMAR, and MGMAR.}
\label{tab:test_score}
\setlength{\tabcolsep}{15pt}
\renewcommand{\arraystretch}{1.3}
\begin{tabular}{lccccc}
Ranking			& 1             & 2             & 3             & NMAR & Proposed      \\ \hline
Final score     & 0.96          & 0.98          & 0.99          & 1.82 & \textbf{0.89} \\ \hline
CTN				& 0.74          & 0.81          & 0.82          & 2.03 & \textbf{0.72} \\
Noise           & \textbf{0.01} & 0.19          & \textbf{0.01} & 1.06 & \textbf{0.01} \\
Sharpness       & 0.72          & \textbf{0.52} & 0.54          & 0.58 & 0.73          \\
Streak          & \textbf{1.29} & 1.65          & 1.48          & 1.99 & 1.33          \\
SSIM            & 0.75          & 0.95          & 0.98          & 2.03 & \textbf{0.72} \\
Metal integrity & 1.46          & 1.09          & \textbf{0.96} & 2.06 & 1.13          \\
Bone integrity  & \textbf{0.89} & 0.99          & 1.09          & 2.05 & 0.93          \\
PBR          	& 1.79          & 1.62          & 2.03          & 2.76 & \textbf{1.55}
\end{tabular}
\renewcommand{\arraystretch}{1.0}
\end{table}

\begin{figure}[ht]
    \centering
    \includegraphics[width=1.0\textwidth]{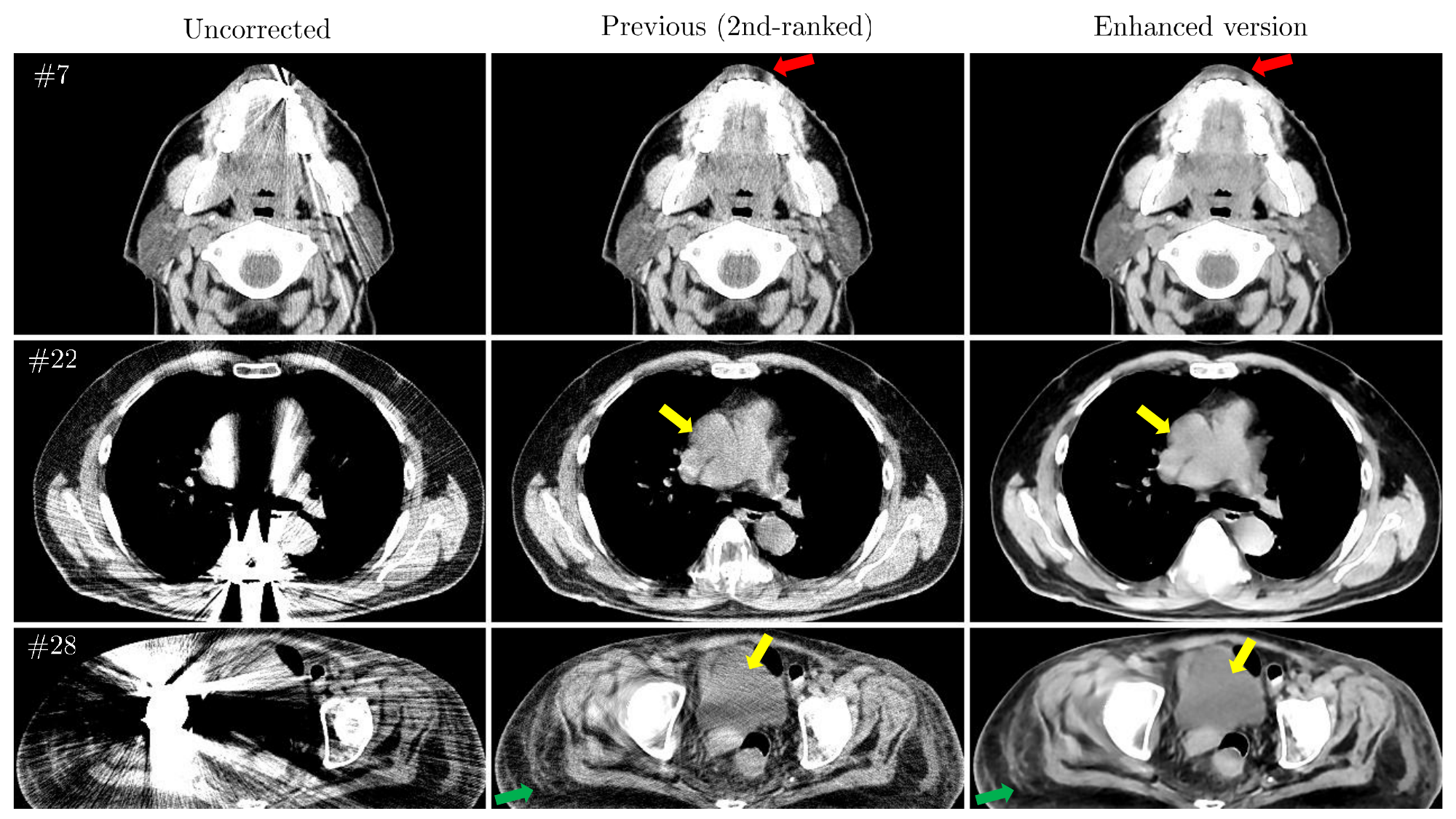}
	\caption{Visual comparison of corrected images obtained by MGMAR and an earlier version of our method (ranked second in the AAPM-MAR challenge) for head (first row), thorax (second row), and pelvis (third row) test cases (WW/WL \(=200/0\) HU).}
    \label{fig:test-results}
\end{figure}

Fig.~\ref{fig:test-results} compares corrected images obtained by MGMAR and an earlier version of our method on representative test cases, including head, thorax, and pelvis.
Compared to the previous version, MGMAR more effectively suppresses streaking and shadowing artifacts near metallic objects (red arrows) and reduces noise in homogeneous tissue regions (yellow arrows). However, it attenuates fine structural details, leading to reduced sharpness in the reconstructed image (green arrows). The corresponding ground-truth images are visualized in the benchmark paper \cite{Haneda2025}. For a qualitative comparison, we adopt the same window setting as used in the benchmark paper. The final reconstructed CT images obtained by MGMAR (raw files) for all test cases, along with their corresponding evaluation metric scores, are provided in the supplementary materials.

\section{Discussion and Conclusion}
In this work, we present MGMAR, a metal-guided MAR framework that explicitly exploits metal-related information throughout the pipeline. Specifically, we design an artifact-aware latent conditioning strategy for prior-image generation. This design enables the latent field (i.e., \(\z=E_{\bpsi}(\bmu)\)) to capture metal-dependent global artifact patterns (e.g., streaking and shadowing), thereby improving the stability and fidelity of the conditioned INR under severe metal corruption (Fig.~\ref{fig:MA_effects}).  In addition, we incorporate explicit metal information in the residual correction stage by conditioning the metal mask into the residual network through an AdaIN-based modulation mechanism. This metal conditioning encourages the residual network to focus its capacity on metal-dependent secondary artifacts induced by projection completion, while reducing unintended distortions in non-metal anatomical regions (Fig.~\ref{fig:res_effects}).

We also propose a data-driven weight initialization strategy for INR-based prior-image reconstruction. Instead of relying on random initialization, we embed data-driven prior knowledge directly into the MLP parameter space by pretraining a conditioned INR on paired metal-corrupted and artifact-free CT images. This prior-embedded initialization guides the measurement-specific refinement toward a more reliable solution manifold under severely limited metal-unaffected measurements, while reducing sensitivity to random initialization and accelerating convergence to a high-quality prior image (Fig.~\ref{fig:prior_image_generation}). 

Related to prior-knowledge embedding, Shen \etal~\cite{Shen2022} leverage an additional CT scan of the same patient as prior information, whose pretrained weights are used to initialize the MLP for sparse-view CT reconstruction. While this strategy can improve reconstruction quality, its applicability may be limited in general clinical workflows where sequential CT scans of the same patient are not routinely available. Meta-learning frameworks embed prior knowledge into the MLP by pretraining on samples from the target domain. Such strategies have been explored for neural radiance fields~\cite{Tancik2021} and CBCT reconstruction~\cite{Shin2025}. Compared to random initialization, meta-learned initializations can improve optimization stability and reduce training time; however, recovering fine anatomical details may still require substantial computation, and the resulting reconstructions can remain suboptimal under severe metal corruption (Fig.~\ref{fig:prior_image_generation}).

The validation results demonstrate that each component of the proposed pipeline contributes consistently to artifact suppression and image-quality improvement (Table~\ref{tab:validation}) in terms of RMSE, PSNR, and SSIM. The prior-image generation stage provides a substantial reduction in reconstruction error, indicating that the conditioned INR yields an effective estimate even under severe metal corruption. The subsequent NMAR step further improves the reconstruction by restricting modifications to the metal-trace region, which helps preserve structural sharpness that is not fully recovered in the INR-based prior image. Finally, the residual correction stage provides an additional improvement by removing secondary artifacts around metallic objects and reducing additional streaking artifacts, leading to the best quantitative performance on the validation set. On the clinical test set, MGMAR outperforms the baseline methods across multiple clinically relevant metrics, including CTN accuracy, noise, SSIM, and PBR influence, and achieves the best final benchmark score on the AAPM-MAR benchmark (Table~\ref{tab:test_score}).

The proposed data-driven initialization relies on paired training datasets with artifact-free references, which may be difficult to acquire in real clinical settings. In practice, vendor-specific, physics-based numerical simulation pipelines can be used to generate realistic training data, thereby mitigating the domain gap between training and clinical domains~\cite{Park2018,Park2022,Zhang2018a}. In addition, although the proposed data-driven initialization significantly reduces the computational burden compared to random and meta initializations, the overall runtime of the proposed method is still approximately \(50\)~s per case, with the majority of the time (\(41\)~s) spent on the measurement-specific refinement. Future work will therefore focus on further accelerating the refinement stage, for example through more efficient optimization strategies, and on extending the proposed framework to real clinical workflows and more complex 3D cone-beam CT geometries.

\section*{Acknowledgment}
H. S. Park and K. Jeon were supported by the National Institute for Mathematical Sciences (NIMS) grant funded by the Korean government (No. NIMS-B26A10000). H. S. Park and K. Jeon were partially supported by the National Research Foundation of Korea(NRF) grant funded by the Korea government(MSIT) (No. RS-2024-00338419).


\end{document}